\documentclass[a4paper]{article}
\usepackage{amsfonts,amsmath,amssymb,amscd}
\newcommand{\be}{\begin{equation}\label}
\newcommand{\ee}{\end{equation}}
\newcommand{\prt}{\partial}
\newcommand{\p}{\prime}

\begin{document}

\title{Double Gauge Invariance and \\Covariantly-Constant Vector Fields in Weyl Geometry} 
\maketitle
\date{~}

\vskip 1cm
\begin{center}
{\large Vladimir~V~Kassandrov\footnote{Institute of Gravitation and Cosmology, Russian Peoples' Friendship University, Moscow, Russia, E-mail: vkassan@sci.pfu.edu.ru} 
and Joseph~A~Rizcallah\footnote{School of Education, Lebanese University, Beirut, Lebanon, E-mail: joeriz68@gmail.com}}
\end{center}


\section{Introduction}
\label{intro}
Weyl unified theory of gravitation and electromagnetism~\cite{weyl19}  
remains one of the most elegant geometrodynamical constructions 
in theoretical physics. Though it is generally accepted that it 
failed to properly describe the observed structure of gravielectromagnetic 
phenomena, the concept of gauge invariance it brought along was inherited by 
quantum theory, where it took the form of a fundamental principle 
governing the structure of all physical interactions. 

As it is well known~\cite{weyl19,rosen82}, Weyl manifolds are defined by the assumption that, contrary to Riemann geometry where the length $l$ is preserved, the parallel transport of a 4-vector alters its length by an amount $dl$ proportional to the length $l$ itself, 
\be{transfer}
dl \propto -l A_\mu dx^\mu. 
\ee
In this way, the \emph{electromagnetic 
4-potential} enters the theory through the 1-form ${\bf A}=A_\mu(x)dx^\mu$. 
Indeed, definition (\ref{transfer}) is equivalent to the following condition 
for the metric tensor $g_{\mu\nu}$ of a Weyl manifold:
\be{metric}
\nabla_\rho g_{\mu\nu} = -A_\rho g_{\mu\nu},
\ee
which is invariant under a \emph{conformal-gauge transformation} 
\be{conform}
g_{\mu\nu} \mapsto \exp{(\lambda)} g_{\mu\nu},~~~A_\mu \mapsto A_\mu -\prt_\mu \lambda,  
\ee
where $\lambda=\lambda(x)$ is an arbitrary differentiable function of coordinates. 
The corresponding 2-form ${\bf F}= d{\bf A}$ is gauge invariant, thus can be identified 
with the electromagnetic field strength. Also recall that condition (\ref{metric}) 
is equivalent to the following expression for the {\it connection} of the Weyl space: 
\be{conn}
\Gamma^\rho_{\mu\nu} = \gamma^\rho_{\mu\nu}+\frac{1}{2}(A_\mu \delta^\rho_\nu+ A_\nu \delta^\rho_\mu- A^\rho g_{\mu\nu}), 
\ee
the first Riemannian term being the Christoffel symbols,
\be{connR}
\gamma^\rho_{\mu\nu}=\frac{1}{2}g^{\rho \alpha}(\prt_\mu g_{\alpha\nu}+
\prt_\nu g_{\alpha\mu}-\prt_\alpha g_{\mu\nu}). 
\ee

We are not in a position to discuss here all the consequences and difficulties of  
Weyl's unified theory, referring the reader instead to, say, 
~\cite{pauli21},~\cite{eddington21},~\cite{Dirac} and historical reviews  in ~\cite{Schmidt,Vizgin} or ~\cite{Filippov}. For our further discussion it is only 
important to point out that, after the establishment of quantum mechanics, 
London~\cite{london27} and Weyl himself~\cite{weyl29} suggested to model electromagnetic interactions through the following well-known substitution of the wave function $\psi(x)$ derivatives 
\be{deriv}
\prt_\mu \psi \mapsto D_\mu\psi, ~~D_\mu:=\prt_\mu - i\frac{e}{\hbar c}A_\mu    
\ee
in the original relativistic field equations for a free particle. This makes the equations 
\emph{gauge invariant} under the joint transformations of electromagnetic 
potentials and the phase of the wave function of the well-known form 

\be{gauge}
A_\mu \mapsto A_\mu + \prt_\mu \theta,~~\psi \mapsto \psi \exp{(i\frac{e}{\hbar c}\theta)},  
\ee
where $\theta=\theta(x)$ is a new gauge function independent of the 
previous one $\lambda(x)$ associated with the above conformal-gauge group 
of transformations (\ref{conform}).

The  idea  of gauge invariance, 
especially its non-Abelian generalization, became a cornerstone of 
quantum field theory and the Standard Model, in particular. However, the construction was inherited from and motivated by the existence of the conformal-gauge group (\ref{conform}) 
of Weyl geometry. The latter, in the process of development of quantum 
theory, has been almost forgotten and, at present, appears to have dropped out from 
mainstream theoretical physics (see, however,~\cite{Pushkin3,GorbPushkin,Rabin}). 
Nonetheless, as we demonstrate in the present paper (Sect. 2), the two groups can, 
rather naturally, coexist giving rise to \emph{double gauge invariance}. 
Some possible consequences of this remarkable fact for quantum theory can then be gleaned. 

The rest of the paper is organized as follows: in Sect. 3 we study the properties of vector fields \emph{covariantly constant} with respect to the Weyl connection (\ref{conn}). The defining equations of such fields alone allow us to pin down the Weyl non-metricity vector $A_\mu(x)$, provided the metric is fixed. In Sect. 4, assuming a Minkowski metric background we find that the non-metricity vector $A_\mu(x)$ reproduces the Lienard-Wiechert electromagnetic ansatz. But, in contrast to classical electrodynamics, the electric charge turns out to be fixed in magnitude, with different signs corresponding to either retarded or advanced fields. We conclude in Sect. 5 with a discussion and some remarks on the significance of the obtained results.

\section{Double gauge invariance in Weyl space}
\label{sec:1}
Consider first a real-valued scalar field $\Psi(x)$ on
the Weyl manifold background. There exist two fundamental invariant differential 
equations for it defined by the {\it Beltrami operators}, namely, the 
{\it eikonal equation}
\be{eik}
g^{\mu\nu} \prt_\mu \Psi \prt_\nu \Psi = 0, 
\ee
and the linear {\it wave equation}                                             
\be{wave}
g^{\mu\nu}\nabla_\mu \prt_\nu \Psi = 0,
\ee
where $\nabla_\mu$ designates the covariant derivative of a vector field 
(particularly, of the gradient field $\prt_\mu \Psi$) with respect to the Weyl 
connection (\ref{conn}). Making use of the decomposition (\ref{conn}) into 
Riemannian (\ref{connR}) and purely Weyl parts, one can represent the wave 
equation (\ref{wave}) in the following form:

\be{waveR}
\square \Psi = -A^\mu \prt_\mu \Psi,
\ee
where $\square \Psi$ stands for the familiar d'Alembertian in Riemannian space,  

\be{dalambert}
\square \Psi:=\frac{1}{\sqrt{-g}}\prt_\mu\left( \sqrt{-g}g^{\mu\nu}\prt_\nu \Psi\right),
\ee
$g$ being the determinant of the metric tensor. 

Both fundamental equations are obviously conformal-gauge invariant with 
respect to the Weyl group of transformations (\ref{conform}).
Moreover, the eikonal equation (\ref{eik}) possesses an additional 
functional invariance (see, e.g.,~\cite{Eik}): if $\Psi(x)$ is a solution of 
(\ref{eik}), then, for an arbitrary differentiable function $\Phi$, 
$\Phi(\Psi(x))$ is also a solution of (\ref{eik}). Note that this 
invariance does not involve the particular form of Weyl connection and is 
valid in any metric space. 

Under the transformation $\Psi \mapsto \Phi(\Psi)$ the gradient field 
$\prt_\mu \Psi$ gets an additional factor 
\be{phase}
\prt_\mu \Psi \mapsto \Phi^\p \prt_\mu \Psi, ~~~\Phi^\p:=d\Phi / d\Psi. 
\ee
It is now easy to see that if this transformation 
is accompanied by the following gradient transformation of potentials 
\be{grad}
A_\mu \mapsto A_\mu - \prt_\mu \ln{\Phi^\p},
\ee
the wave equation (\ref{waveR}) also preserves its form!  

Unlike the $SU(1)$ gauge group (\ref{gauge}) of quantum theory 
that transforms the {\it phase} of the wave function, the 
additional gauge invariance in Weyl space (\ref{phase},\ref{grad}) deals with 
real-valued transformations which change the magnitude of the scalar field $\Psi$.  
However, if one considers a \emph{complexification} of the 
scalar field $\Psi$, then the gauge parameter $\Phi^\p(\Psi(x))$ in (\ref{phase}) 
will be also complex-valued, and the gauge transformations  will 
act on the phase of $\Psi$ while its modulus can be kept fixed through 
a normalization procedure.

Remarkably, even on the Minkowski metric background, making use of the 
additional gauge invariance of the wave equation (\ref{waveR}), we can 
\emph{geometrically} introduce a new form of interaction of 
a scalar massless field $\Psi(x)$ with an external electromagnetic field 
identified with a ``Minkowski residue'' 
of the Weyl non-metricity vector $A_\mu(x)$.

Of course, the possibility to introduce an alternative, geometrically 
justified form of interaction also preserving the {\it local gauge invariance} 
can be much interesting for 
quantum field theory. Note, however, that the presence of the ordinary mass term 
in (\ref{waveR}) spoils the new gauge invariance, if only one disregards the concept of a \emph{conformal mass} (see, e.g.,~\cite{Barut,Pervushin}). 
The solution of (\ref{waveR}) in a Coulomb external field is presented and discussed below. 

\section{Covariantly-constant fields in Weyl space} 
\label{sec:2}
Let us now consider the structure of covariantly constant vector fields (CCVF), 
i.e. those mapping into themselves in the process of parallel transport with 
respect to the full Weyl connection (\ref{conn}). In general, CCVF is a 
remarkable mathematical and physical object  which is closely connected with 
the structure of holonomy groups~\cite{Hall,Hall2}, shear-free null congruences 
of rays, twistors and biquaternionic analyticity~\cite{GR95,KasJos}, etc. Moreover, overdetermined equations, akin to those of CCVF and their integrability conditions, appear naturally in other fields, such as the method of Ernst potential in  General Relativity~\cite{HS} and the representation of null curvature in nonlinear field theory~\cite{FT}. In the context of Weyl unified theory, bringing the CCVF equations to the forefront as the principal object of physical dynamics could help evade the non-uniqueness of scales issue, which  was first pointed out by Einstein~\cite{Einstein} and which presented a major difficulty to the old theory. 

The defining equations of the CCVF   
\be{ccvf}
\nabla_\mu K_\nu = \prt_\mu K_\nu - \Gamma^\rho_{\mu\nu}K_\rho =0,
\ee
with $\Gamma^\rho_{\mu\nu}$ being the full connection (\ref{conn}), are 
overdetermined (16 equations for 4 unknown components of the CCVF $K_\mu(x)$) 
and do not admit solutions in an arbitrary Weyl space. The very assumption on the existence of a CCVF immediately imposes rigid restrictions on the structure and holonomy type of Weyl connection~\cite{Hall3}. 
 
Moreover, even in the limit of Minkowski metric, the equations fully determine the 
Weyl non-metricity vector $A_\mu(x)$, hence the associated 
electromagnetic field. On the other hand, to fix the metric and Weyl non-metricity vector both together one could supplement  (\ref{ccvf}) by additional  equations. The latter should then be conformally invariant though consistent with the Einstein equations in General Relativity (at least in the Newtonian limit); this issue is discussed in the conclusion. 

Recall also that on a purely Riemannian background any CCVF $L_\mu(x)$, such that 
$\tilde \nabla_\mu L_\nu =0$, 
with $\tilde \nabla$ being the covariant derivative w.r.t. the Levi-Civita connection (\ref{connR}),  is, of course, a {\it Killing vector field}, $\tilde \nabla_\mu L_\nu+\tilde \nabla_\nu L_\mu =0$.  So one has no CCVF in spaces without symmetries, while in product spaces all CCVF are rather trivial. In contrast, on the Weyl space background the CCVF Eqs. (\ref{ccvf}) have no direct relationship to the symmetry of the metric. Therefore, a {\it nontrivial} CCVF can exist in maximally symmetric spaces, such as  Minkowski (see below), as well as in spaces with no symmetries.

Moreover, as 
it was shown back in~\cite{Kasan,Acta}, even in the simplest case of Weyl's space 
with connection (\ref{conn}) and constant Minkowski metric, equations (\ref{ccvf}) 
possess an array of remarkable features. In particular, due to their 
overdeterminacy, they yield the {\it self-quantization} of the 
electric charge of field singularities. The general principles of CCVF 
geometrodynamics were formulated in subsequent papers~\cite{AD,Master,GR95}, 
where other examples of connections admitting an electromagnetic 
field-theoretical interpretation were also considered.

Let us proceed with the investigation of the CCVF system of Eqs. (\ref{ccvf}),(\ref{conn}). The integrability conditions are obtained by equating to zero the commutator 
\be{int-ccvf}
\nabla_{[\mu} \nabla_{\nu]}K_\rho = R^\alpha_{\rho[\mu\nu]}K_\alpha =0,
\ee    
where the curvature tensor, corresponding to connection (\ref{conn}), is given by
\be{cur-weyl}
R_{\alpha\rho[\mu\nu]}=P_{[\alpha\rho][\mu\nu]}+\frac{1}{2}g_{\alpha\rho}F_{[\mu\nu]},
\ee   
with $F_{[\mu\nu]}=\prt_{[\mu} A_{\nu]}$ regarded as the electromagnetic 
field strength tensor. By its algebraic properties the tensor 
$P_{[\alpha\rho][\mu\nu]}$ is analogous to the Riemann tensor; in particular, 
it is skew symmetric in both pairs of indices. So from 
(\ref{int-ccvf},\ref{cur-weyl}) we readily find
\be{norm}
\left\|K\right\|F_{[\mu\nu]}=0,	
\end{equation}
where $\left\|K\right\|=g^{\alpha\beta}K_\alpha K_\beta$ is the (squared) norm of the 
CCVF $K_\alpha(x)$. It thus follows~\cite{Hall3} that only null CCVF can exist
on manifolds with a non-trivial Weyl 2-form. Physically, this means that nontrivial electromagnetic fields can be  
associated only with \emph{null} CCVF; the latter preserve their components and 
zero length under parallel transport along any closed path on the Weyl space.  Non-null CCVF correspond to the 
pure gauge form of 4-potentials $A_\mu = \prt_\mu \Sigma(x)$; for them the parallel transport is identically 
integrable. However, we shall not consider the non-null CCVF below.

Using now the symmetry of connection (\ref{conn}), from (\ref{ccvf})  
we find
$$\prt_{[\mu} K_{\nu]} =0,$$
so that any CCVF is locally a 4-gradient
\be{grad2}
K_\mu=\partial_\mu\Psi,
\ee
and, by the null property of CCVF (\ref{norm}), the generating function 
$\Psi(x)$ must satisfy the \emph{eikonal equation} (\ref{eik}).

Substituting (\ref{grad2}) into (\ref{ccvf}), one reduces the latter into
\be{ccvfN}
\nabla_\mu \prt_\nu \Psi = \prt_\mu \prt_\nu \Psi- 
\Gamma^\rho_{\mu\nu}\prt_\rho \Psi=0.
\ee
Upon contracting the indices, the wave equation (\ref{waveR}) follows. Therefore, the generating function $\Psi(x)$ of any null CCVF 
{\it must satisfy both fundamental constraints}, the eikonal and wave equations. 

Like the latter, the complete set of equations for null CCVF 
(\ref{ccvfN}) is obviously invariant under transformations of 
the Weyl conformal-gauge group (\ref{conform}). As to the additional group of 
the joint gauge transformations (\ref{phase}),(\ref{grad}), it is \emph{not} a 
symmetry of ({\ref{ccvfN}). Nonetheless, taking into account that the function 
$\Psi(x)$ satisfies the eikonal equation (\ref{eik}), one obtains instead the 
following joint transformations of $\Psi$ and potentials $A_\mu$:
\be{gauge2}
\Psi \mapsto \Phi(\Psi),~~~(\prt_\mu \Psi \mapsto \Phi^\p \prt_\mu \Psi),~~~
A_\mu \mapsto A_\mu + \prt_\mu \ln{\Phi^\p},
\ee
which \emph{leave the complete set of equations (\ref{ccvfN}) for null CCVF 
form-invariant} (pay attention to the opposite signs of the gradient terms for the 
transformed potentials in (\ref{grad}) and (\ref{gauge2})). Note that this 
form-invariance does not contradict that represented by (\ref{phase}),(\ref{grad}) 
of the wave equation (\ref{waveR}). Indeed, if the generating function 
$\Psi$ in the latter does, in addition, satisfy the eikonal equation (\ref{eik}), 
its group of symmetry becomes wider:
\be{gaugeW} 
\Psi \mapsto \Phi(\Psi),~~~(\prt_\mu \Psi \mapsto \Phi^\p \prt_\mu \Psi),~~~
A_\mu \mapsto A_\mu + \kappa\prt_\mu \ln{\Phi^\p},
\ee
with $\kappa$ being an \emph{arbitrary} number; in particular, the case 
$\kappa=0$ is allowed in which one does not transform the potentials $A_\mu$
at all but only the generating function $\Psi(x)$ (obeying necessarily 
the eikonal equation!). All these results can be confirmed by simple 
direct calculations.

Let us now restrict ourselves to the case of constant Minkowski metric 
$g_{\mu\nu}=\eta_{\mu\nu}=diag\{-1,1,1,1\}$ (or, more generally, to spaces \emph{conformally flat in the Riemann sense}). 
In this case the Levi-Civita 
part (\ref{connR}) of the full connection (\ref{conn}) vanishes, and the set of 
equations for null CCVF (\ref{ccvfN}) simplifies to 
\be{ccvfNMink}
\prt_\mu \prt_\nu \Psi= \frac{1}{2}(A_\mu \prt_\nu \Psi + A_\nu\prt_\mu \Psi -
\eta_{\mu\nu} A^\rho \prt_\rho \Psi). 
\ee

Consider next the solution of (\ref{ccvfNMink}) corresponding to the \emph{spherical 
wave solution} of the eikonal equation (\ref{eik}), 
\be{sphsol}
\Psi=r+\epsilon t,~~~ r:=\sqrt{x^a x_a},~~a=1,2,3,
\ee
where the quantity $\epsilon=\pm1$ stands for the speed of light in natural 
units. It's straightforward to calculate the components of the null CCVF (up to 
a multiplicative constant) and 
electromagnetic  potential corresponding to this solution; the result reads 
\be{sol}
K_0=\epsilon=\pm1,K_a=x^a/r;\ \ A_0=\epsilon/r, A_a=-x^a/r^2.
\ee
Hence, for the associated  electromagnetic field one has the Coulomb-like ansatz 
with a fixed (equal to unity) value of the electric charge 
$\epsilon = \pm1$ (the presence of the magnetic potential, being a pure gauge, does 
not alter the field strength). It's interesting that 
the Coulomb solution is generated here 
by the spherical wave distribution for the function $\Psi$ while the 
``quantization'' of the electric charge comes about as a consequence of the 
fixed value of the wavefront's speed of propagation. Note, in passing, that the plane wave solution of (\ref{eik}) $$\Psi= x + \epsilon t$$ has trivial associated electromagnetic fields.

\section{Weyl geometric origin of the Lienard-Wiechert electromagnetic fields} 
\label{sec:3}
Consider now the CCVF potentials produced by a point-like charge moving along 
an arbitrary worldline $\xi^\mu(\tau)$ in Weyl space with constant Minkowski 
metric $\eta_{\mu\nu}$. For a space-time point $x^\mu$ the light cone equation 
\be{retard}
X_\mu X^\mu = 0, ~~~~X^\mu: = x^\mu -\xi^\mu(\tau)
\ee
implicitly defines the \emph{field of retarded/advanced time-like parameter} 
$\tau=\tau(x)$ which is known to obey the eikonal equation 
\be{eik2}
\eta^{\mu\nu}\prt_\mu \tau \prt_\nu \tau = 0.
\ee
Indeed, differentiating (\ref{retard}) with respect to $x^\mu$, 
one obtains
\be{prtau}
K_\mu := \prt_\mu \tau = \frac{X_\mu}{P}
\ee
provided the quantity $P:=X_\lambda \xi^{\p\lambda}$ (the so-called 
``retarded distance'') does not turn to zero. 
Taking then the Lorentz norm of the gradient field (\ref{prtau}), one 
obtains in view of (\ref{retard}) the eikonal equation (\ref{eik2}).

Let us use the field $\tau(x)$ as a generating function $\Psi(x)$ for 
null CCVF. It is noteworthy that the symmetry group 
of gauge transformations (\ref{gauge2}) 
corresponds then to a symmetry of the light cone equation with respect to 
\emph{reparametrization} of the time-like parameter $\tau \mapsto f(\tau)$, with 
a monotonic function $f(\tau)$. Note also that the null CCVF $K_\mu(x)$ as 
given by (\ref{prtau}) is the field tangent to a shear-free and rotation-free 
congruence of null ``rays'' divergent from the point-like  source on a  
worldline.

Making now use of (\ref{prtau}) to calculate  
the second order derivatives, one finds (compare, say, with~\cite{Bonnor})  
\be{secderiv}
\prt_\mu\prt_\nu \tau = \frac{1}{P}(\eta_{\mu\nu} - (\xi_\mu^\p K_\nu +
\xi_\nu^\p K_\mu) + K_\mu K_\nu 
(\xi_\lambda^\p \xi^{\p\lambda} -X_\lambda \xi^{\p\p\lambda})),
\ee
with $\p$ designating differentiation with respect to $\tau$. Now Eqs.  
(\ref{ccvfNMink}) should define the potentials $A_\mu(x)$. To simplify calculations, 
let us first satisfy the contracted Eq. (\ref{waveR}). Making use 
of (\ref{secderiv}) and (\ref{eik2}), one finds for the d'Alembertian (\ref{dalambert})
\be{dalamb}   
\square \tau = \frac{2}{P},
\ee
and the contracted Eq. (\ref{waveR}) takes the form
\be{contr}
A^\mu X_\mu =-2,
\ee
so for the potentials $A_\mu(x)$ one can assume the following Lorentz 
covariant ansatz:
\be{ansatz}
A_\mu = -\frac{2}{P} \xi_\mu^\p + \beta(x) K_\mu,
\ee
with the scalar function $\beta(x)$ to be determined from the full set of 
CCVF defining Eqs. (\ref{ccvfNMink}). 
Substituting in the latter the ansatz for potentials (\ref{ansatz}), 
one finally obtains 
\be{beta}
\beta = 
\frac{1}{P}(\xi_\lambda^\p \xi^{\p\lambda} -X_\lambda \xi^{\p\p\lambda}),
\ee
so that the potential $A_\mu(x)$ is given by (\ref{ansatz}) together with 
(\ref{beta}), and the components of the corresponding null CCVF are found from 
(\ref{prtau}), with $\tau(x)$ being implicitly defined by (\ref{retard}). 

Moreover, it is possible to separate a full 4-gradient from the expression for 
potentials (\ref{ansatz},\ref{beta}) through the representation of the second term 
in (\ref{ansatz}) in the following equivalent form:
\be{equiv}
\beta(x) K_\mu \equiv \frac{1}{P}\xi_\mu^\p -\prt_\mu \ln P. 
\ee
Combining the terms in (\ref{ansatz}),(\ref{equiv}) and discarding the 
4-gradient term which does not contribute to the electromagnetic 
field strength, we obtain for the equivalent potentials $\tilde A_\mu(x)$ 
the following expression reproducing the familiar Lienard-Wiechert form:
\be{lienard}
\tilde A_\mu = -\frac{\xi_\mu^\p}{P} \equiv \epsilon\frac{u_\mu}{X_\lambda u^\lambda},
\ee
where $u_\mu:=\xi_\mu^\p(\tau)$ is the 4-velocity vector, and for \emph{retarded} 
time solution for which $X^0 = x^0-\xi^0 >0$ the dimensionless electric charge 
$\epsilon=-1$ whereas the opposite value $\epsilon=+1$ corresponds to \emph{advanced} potentials. 
In either case the value of charge is fixed, ``self-quantized'' (the constant 
$\epsilon$ is the same for an arbitrary worldline of a point-like source). 
For a charge at rest the solution obviously turns into the previously found  
Coulomb-like one (\ref{sphsol}),(\ref{sol}).                                  

Some remarks are to the point here. In fact, we have demonstrated that the 
principal solution to Maxwell equations, the Lienard-Wiechert field, 
naturally emerges within the framework of CCVF in Weyl geometry. This gives the field part 
of classical electrodynamics a transparent geometrical meaning. 
However, one has two remarkable distinctions between the canonical and the presented 
approaches. The first is related to the already mentioned ``self-quantization'' 
effect: only sources with a unit (\emph{elementary}) charge can produce 
the electromagnetic field associated with (null) CCVF on the Weyl geometric background. 

The second peculiarity of the above presented results is the fundamental 
distinction between point-like sources with opposite signs of the electric 
charge, one of which generates the \emph{retarded} 
Lienard-Wiechert field while the other the \emph{advanced} one. This  
remarkable result allows one to explicitly connect the 
\emph{particle-antiparticle asymmetry with causality and the ``arrow of time''}. The situation strongly resembles Feynman's representation of a ``positron as a moving backward-in-time 
electron''; however, here, this arises as a direct consequence of Weyl geometry and CCVF structure rather than being a useful hypothesis! Needless to say, classical electrodynamics admits no analog of this feature. 

\section {Conclusion}
\label{sec:4}

In the paper, some remarkable and  rather unexpected results related to geometrophysical objects defined on the Weyl space have been obtained, among which are: 
\vskip0.1cm
\noindent

1) the {\it double gauge invariance} of the canonical fundamental equations on the Weyl manifold, the wave-like equation in particular, and the alternate, purely geometric realization of the {\it local gauge invariance principle}, 

2) the close connection of the additional gauge symmetry group and the group of {\it reparametrization} of the equations of  a point-like source  worldline, 

3) the {\it ``self-quantization''} of the electric charge value for the ordinary Lienard-Wiechert field, and the close relation of the phenomenon with the universality of the speed of light,

4) the explicit connection between the ``arrow of time'' (represented in the distinction of retarded/advanced fields) and the particle/antiparticle asymmetry (represented in different signs of charges of corresponding sources); this situation gives geometric ``flavor" to the well-known Feynman's construction for positrons.

\vskip0.1cm
\noindent

The considered model of a CCVF-geometrization of the electromagnetic field offers an overdetermined non-Lagrangian system of nonlinear partial differential 
equations rather rarely used in theoretical physics. On the Minkowski metric 
background, the CCVF system completely determines the structure of both the CCVF and 
electromagnetic field. For the fundamental solution, the CCVF gives rise to a 
shear-free null congruence of rays emanating from a point-like source moving 
along an arbitrary worldline while the electromagnetic potential corresponds to the canonical 
Lienard-Wiechert ansatz. On the one hand, the latter provides a transparent 
geometric interpretation of the electromagnetic field and, on the other, satisfies Maxwell equations  which, in a sense, are \emph{replaced} by the equations of CCVF. 

Indeed, the free Maxwell equations themselves do not follow from  
the CCVF system. Moreover, the considered system violates the principle of 
superposition, and in order to find a class of solutions corresponding to 
a {\it number} of moving charges, one could make use, say, of the Wheeler-Feynman 
``one electron Universe'' conjecture on the ensemble of identical point-like 
particles on a unique worldline (an attempt to realize this conjecture had been 
recently undertaken in~\cite{Ildus}). 

On the other hand, under an 
admissible scale change of the CCVF $K_\mu(x) \Rightarrow CK_\mu(x),~C=const$ 
the connection and the corresponding electromagnetic potentials remain unaltered. 
Along with the fact that the CCVF system is overdetermined this indicates that 
some dynamic restrictions exist on the  electromagnetic characteristics of  
solutions, first and foremost, the electric charge. So it comes as no surprise that the magnitude of electric charge for the fundamental Lienard-Wiechert solution turns out to be unity, \emph{elementary}. This result strongly correlates  with Einstein's views on the possible fundamental role of {\it overdetermined} systems of partial differential equation as the generators of quantization conditions~\cite{Ein1,Ein2}, in our case, for the electric charge.

In another respect, the offered CCVF system looks \emph{underdetermined}, being 
insufficient, by itself, to determine the conformal class of the metric. So one 
has to supplement the system with an additional dynamical restriction  
following, probably, from a (conformal-invariant) Lagrangian. In this 
case, the CCVF system can enter the theory through, say, a Lagrangian 
multiplier procedure. However, the choice of a suitable Lagrangian in the 
framework of Weyl geometry is rather a difficult and yet unsolved problem 
(see, e.g.,~\cite{Schmidt,Schmidt2}). Probably, the problem can be 
simplified if one requires the invariance of the full Lagrangian under the above obtained 
\emph{double gauge symmetry}.

In an alternative approach, one can consider the joint system of the CCVF equations and the so-called "geometrodynamical equations'' of Gorbatenko-Pushkin (see, e.g.,~\cite{Pushkin3} and references therein)
\be{pushkin}
 R_{(\mu\nu)}^*=0, 
\ee
with symmetric part of the Ricci tensor defined on the Weyl manifold in the l.h.s.  These conformally invariant  equations seem to be compatible with those of CCVF (note that from (\ref{pushkin}) free Maxwell equations follow directly~\cite{Pushkin3}).  
Together with (\ref{pushkin}), the CCVF equations (\ref{ccvf}) form a self-consistent 
dynamical {\it non-Lagrangian} system capable of determination of gravitational and electromagnetic fields both together. Note also that, in contrast to conformally invariant equations of the "fourth order gravity", equations 
(\ref{pushkin}) are of the second order in metrics and can reproduce the canonical effeccts of GR in the Newtonian limit. 
   
Finally, we conclude that although the well-known difficulties encountered by Weyl's old unified theory (absence of a suitable Lagrangian, ``unphysical'' structure of geodesics, etc.) still cannot be considered completely
resolved within the presented approach, they  just don't look so despairing in view of the results obtained so far, allowing one to surmise the future \emph{revival} of Weyl's elegant theory.
  
\vskip 0.2cm 
\noindent
{\bf Acknowledgements} We would like to thank Profs. D.V. Alexeevski, A.Ya. Burinskii, B.N. Frolov and A.S. Rabinowitch for friendly discussions and valuable comments. The authors are also indebted to the referees for critical remarks and advices which 
helped to improve the paper and stimulated subsequent work. One of the authors (V. K.) wants to express his particular gratitude to Prof. E.T. Newman for the long-term support and kind attention.



\end{document}